\documentclass[12pt]{amsart}
\usepackage{amsmath}
\usepackage{amsthm}
\usepackage{amsfonts}
\usepackage{graphicx}
\usepackage{amssymb}% for \sphericalangle

\begin{document}

\title{Square--roots and Lax--pairs for supersymmetrizable systems}
\author[Jens Hoppe]{JENS HOPPE}
%\date{June 14, 2020}
\address{Braunschweig University}
\email{jens.r.hoppe@gmail.com}

\begin{abstract}
Several examples are given illustrating the (presumably rather general) fact that bosonic Hamiltonians that are supersymmetrizable automatically possess Lax--pairs, and square--roots.
\end{abstract}

\maketitle
%%%%%%%%%%%%%%%%%%%%%%%%%%%%%%%%%%%%%%%%%%%%%%%%%%%%%%%%%%%%%%%%%%%%%%%%%%%%%
%PAGE 1 PAGE 1 PAGE 1 PAGE 1 PAGE 1 PAGE 1 PAGE 1 PAGE 1 PAGE 1 PAGE 1 PAGE 1
%%%%%%%%%%%%%%%%%%%%%%%%%%%%%%%%%%%%%%%%%%%%%%%%%%%%%%%%%%%%%%%%%%%%%%%%%%%%%
\noindent
As is well known, the matrix models\cite{1} describing bosonic membranes in $D$--dimensional space--time can be supersymmetrized \cite{2} for $D =$4, 5, 7 and 11, and coincide with quantum-mechanical models \cite{3} that are reductions of Super--Yang--Mills theories. So, as a first class of examples, consider the bosonic Hamiltonians \cite{1}
\begin{equation}\label{eq1}
H = \dfrac{1}{2}\text{Tr}\big( \vec{P}^2 - \sum_{s<t} [X_s, X_t]^2  \big),
\end{equation}
%%%%%%%%%%%%%%%%%%%%%%%%%%%%%%%%%%%%%%%%%%%%%%%%%%%%%%%%%%%%%%%%%%%%%%%%%%%%%
%PAGE 2 PAGE 2 PAGE 2 PAGE 2 PAGE 2 PAGE 2 PAGE 2 PAGE 2 PAGE 2 PAGE 2 PAGE 2
%%%%%%%%%%%%%%%%%%%%%%%%%%%%%%%%%%%%%%%%%%%%%%%%%%%%%%%%%%%%%%%%%%%%%%%%%%%%%
resp.
\begin{equation}\label{eq2}
\begin{split}
L_{\beta} & := \big( p_{ta} \gamma^t_{\beta \alpha} + \dfrac{1}{2} f_{abc} x_{sb}x_{tc}\gamma^{st}_{\beta \alpha} \big) \theta_{\alpha a} \\
 & =: V^{\beta}_{\alpha a} \theta_{\alpha a} \\
 & = \big( P^{\beta}_{\alpha a} + Q^{\beta}_{\alpha a} \big)\theta_{\alpha a}
\end{split}
\end{equation}
where $s, t = 1, \ldots, d := D-2$, (\textit{not} restricted to 2, 3, 5, 9) $\alpha,\, \beta = 1, \ldots, \Sigma$, $abc = 1,\ldots,N^2-1$, $f_{abc}$ real, totally antisymmetric, structure constants of $su(N)$,
\begin{equation}\label{eq3}
\begin{split}
\gamma^s \gamma^t + \gamma^t \gamma^s & = 2 \delta^{st}\mathbf{1}_{\Sigma \times \Sigma},\\
\theta_{\alpha a}\theta_{\beta b} + \theta_{\beta b}\theta_{\alpha a} & = \delta_{\alpha \beta}\delta_{ab}\mathbf{1},
\end{split}
\end{equation}
the $\gamma^{\cdot}_{\cdot \cdot}$ being real constants, while the $\theta_{\alpha a}$ are constant (i.e. in contrast to the supersymmetric models \cite{3} non--dynamical) hermitean matrices (of appropriate size). One easily checks that

\begin{equation}\label{eq4}
\begin{split}
P^{\beta}_{\alpha a} P^{\beta'}_{\alpha a} &= \delta^{\beta \beta'}p_{ta}p_{ta}\\
Q^{\beta}_{\alpha a} Q^{\beta'}_{\alpha a} &= \delta^{\rho \rho'}2V \\
P^{\beta}_{\alpha a}Q^{\beta'}_{\alpha a} + P^{\beta'}_{\alpha a}Q^{\beta}_{\alpha a}
& = 2 \gamma^t_{\beta \beta'} x_{ta}J_a \\
V^{\beta}_{\alpha a} V^{\beta'}_{\alpha a}
& = \delta_{\beta \beta'} 2 H + 2 \gamma^t_{\beta \beta'} x_{ta} (f_{abc} x_{sb} p_{sc})\\
H & = \dfrac{1}{2}(p_{ta}p_{ta}) + \dfrac{1}{4} f_{abc}f_{ab'c'}x_{sb} x_{tc} x_{sb'} x_{tc'}
=: T+V
\end{split}
\end{equation}
%%%%%%%%%%%%%%%%%%%%%%%%%%%%%%%%%%%%%%%%%%%%%%%%%%%%%%%%%%%%%%%%%%%%%%%%%%%%%
%PAGE 3 PAGE 3 PAGE 3 PAGE 3 PAGE 3 PAGE 3 PAGE 3 PAGE 3 PAGE 3 PAGE 3 PAGE 3
%%%%%%%%%%%%%%%%%%%%%%%%%%%%%%%%%%%%%%%%%%%%%%%%%%%%%%%%%%%%%%%%%%%%%%%%%%%%%
Hence (for any $d$)
\begin{equation}\label{eq5}
L_{\beta}L_{\beta'} + L_{\beta'}L_{\beta} = 2\delta_{\beta \beta'} H\mathbf{1} + 2\gamma^t_{\beta \beta'}x_{ta}J_a,
\end{equation}
( in particular, $L^2_{\beta} = H \cdot \mathbf{1}$ in the $su(N)$ invariant sector $J_a = 0$ ) and the classical equations of motion,
$$
\dot{x}_{sb} = p_{sb}, \quad \dot{p}_{sb} = -\dfrac{\partial H}{\partial x_{sb}}
$$
imply
\begin{equation}\label{eq6}
\begin{split}
\dot{Q}^{\beta}_{\alpha a} & = \Omega_{\alpha a, \alpha' a'} P^{\beta}_{\alpha' a'},\\
\dot{P}^{\beta}_{\alpha a} & = \Omega_{\alpha a, \alpha' a'} Q^{\beta}_{\alpha' a'}, \\
\Omega_{\alpha a, \alpha' a'} & := -f_{aba'}\gamma^s_{\alpha \alpha'} x_{sb}
\end{split}
\end{equation}
(hence $\dot{V}^{\beta}_{\alpha a} = \Omega_{\alpha a, \alpha' a'}V^{\beta}_{\alpha' a'}$ ) as well as (for $J_a = 0$) the Lax--pair equation(s)
\begin{equation}\label{eq7}
\dot{L}_{\beta} = \big[ L_{\beta},\, M := \dfrac{1}{2}x_{tc}f_{abc}\gamma^t_{\alpha \beta} \theta_{\alpha a}\theta_{\beta b} \big]
\end{equation}
( resp., even with spectral parameters , $\lambda_{\beta}$, $\dot{L}(\vec{\lambda}) = \big[ L(\vec{\lambda}), M \big]$ for $L(\vec{\lambda}) := \sum\limits_{\beta} \lambda_{\beta} L_{\beta}$;
note however that the usual conserved quantities, from $\text{Tr}\big(L(\vec{\lambda})\big)^k$, seem to all be functionally dependent, due to (\ref{eq5})). A very interesting aspect of this approach (hopefully generalizable to the full Yang--Mills theory) is also(\ref{eq6}),
%%%%%%%%%%%%%%%%%%%%%%%%%%%%%%%%%%%%%%%%%%%%%%%%%%%%%%%%%%%%%%%%%%%%%%%%%%%%%
%PAGE 4 PAGE 4 PAGE 4 PAGE 4 PAGE 4 PAGE 4 PAGE 4 PAGE 4 PAGE 4 PAGE 4 PAGE 4
%%%%%%%%%%%%%%%%%%%%%%%%%%%%%%%%%%%%%%%%%%%%%%%%%%%%%%%%%%%%%%%%%%%%%%%%%%%%%
which e.g. naturally suggests to look for `self--dual' solutions of the from
\begin{equation}\label{eq8}
P^{\beta}_{\alpha a} = A_{\alpha a, \alpha' a'} Q^{\beta}_{\alpha' a'},
\end{equation}
where $A$ (if constant) would have to satisfy $A \Omega A = \Omega$ while naturally antisymmetric, to trivially satisfy the $PQ +QP \cong 0$ relation in (\ref{eq4}).
A simple candidate would be
$ A_{\alpha a, \alpha' a'} = \delta_{a a'} \gamma_{\alpha \alpha'}$
with $\gamma \gamma^s \gamma = \gamma^s$ ; for reduced Yang--Mills theory in Euclidean space, say $\mathbb{R}^4$, one would need the opposite sign and could e.g. take
$ \gamma = -\mathbf{1} \times \epsilon$, if
$\gamma_1 = \sigma_3 \times 1$, $\gamma_2 = \sigma_2 \times \sigma_2$, $\gamma_3 = \sigma_1 \times 1$; and, with $\gamma_{12} = -\sigma_1 \times \epsilon$, $\gamma_{23} = -\sigma_3 \times \epsilon$, $\gamma_{31} = -\epsilon \times 1$ verify that (\ref{eq8}) in that case corresponds to the ordinary self--duality equations,
$p_{ia} = \frac{1}{2}\epsilon_{ijk}f_{abc}x_{jb}x_{kc}$.\\[0.15cm]
%%%%%%%%%%%%%%%%%%%%%%%%%%%%%%%%%%%%%%%%%%%%%%%%%%%%%%%%%%%%%%%%%%%%%%%%%%%%%
%PAGE 5 PAGE 5 PAGE 5 PAGE 5 PAGE 5 PAGE 5 PAGE 5 PAGE 5 PAGE 5 PAGE 5 PAGE 5
%%%%%%%%%%%%%%%%%%%%%%%%%%%%%%%%%%%%%%%%%%%%%%%%%%%%%%%%%%%%%%%%%%%%%%%%%%%%%
Another class of examples is given by ($W = W(x_1,\ldots,x_n)$)
\begin{equation}\label{eq9}
\begin{split}
L_{1} & = \sum^N_{k=1}(f_k p_k - g_k \partial_k W)\\
L_{2} & = \sum(g_k p_k + f_k \partial_k W),
\end{split}
\end{equation}
with hermitean matrices $f_k$ and $g_l$ satisfying standard Clifford anti--commutation relations,
\begin{equation}\label{eq10}
\lbrace f_k, g_l \rbrace = 0, \quad \lbrace f_k, f_l \rbrace = 2 \delta_{kl}\mathbf{1},
\quad \lbrace g_k, g_l \rbrace = 2 \delta_{kl}\mathbf{1}.
\end{equation}
One easily sees that\footnote{Note that the $\partial^2_{kk}W$ terms ( absent in (\ref{eq11}) ) that are often ( e.g. for Calogero--Moser systems ) but not always ( e.g. not for $W = \vec{q}_1 \cdot (\vec{q}_2 \times \vec{q}_3)$, which corresponds to $(\ref{eq1})_{N=2, d=3}$ ) crucial, are a quantum--effect.}
\begin{equation}\label{eq11}
\lbrace L_{\beta}, L_{\beta'} \rbrace = 2 \big( \vec{p}\,^2 + (\vec{\triangledown}W)^2 \big) \mathbf{1}.
\end{equation}
and that the classical equations of motion
($\dot{\vec{x}} = \vec{p}$, $\dot{p}_k = - \vec{\triangledown}W \cdot \partial_k \vec{\triangledown}W$) imply
\begin{equation}\label{eq12}
\dot{L}_{\beta} = \big[ L_{\beta}, M:= - f_k g_l \partial^2_{kl}W \big].
\end{equation}

\vspace{1.5cm}
\noindent
\textbf{Acknowledgement.}
I am grateful to A. Jevicki, D.O'Connor, and T. Turgut for valuable discussions.


\begin{thebibliography}{11111}
\setlength{\baselineskip}{0.9\baselineskip}
\bibitem[1]{1} J.Hoppe , Ph.D. thesis, MIT 1982 http://dspace.mit.edu/handle/1721.1/15717
\bibitem[2]{2} B.deWit, J.Hoppe, H.Nicolai, Nucl.Phys.B320, 1989
\bibitem[3]{3} M.Claudson, M.Halpern, Nucl.Phys.B250, 1985. \\
               R.Flume, Ann.Phys.164, 1985. \\
               M.Baacke, P.Reinicke, V.Rittenberg, J.Math.Phys. 26, 1985.

\end{thebibliography}
\end{document}